\documentclass[article,amsmath,amssymb,superscriptaddress,longbibliography,aps,prl]{revtex4-2}
\usepackage[colorlinks,
linkcolor=blue,
anchorcolor=blue,
citecolor=blue,
]{hyperref}
\usepackage{graphicx}
\usepackage{float}
\begin{document}

\title {Thermomodulated intrinsic Josephson effect in Kagome CsV$_3$Sb$_5$}

\author{Tian Le}
\email{letian\_phy@zju.edu.cn}
\affiliation{Center for Quantum Matter, School of Physics, Zhejiang University, Hangzhou 310058, China}
\affiliation{Key Laboratory for Quantum Materials of Zhejiang Province, Department of Physics, School of Science and Research Center for Industries of the Future, Westlake University, Hangzhou 310030, P. R. China}
\affiliation{Institute for Advanced Study in Physics, Zhejiang University, Hangzhou 310027, China}
\affiliation{Institute of Natural Sciences, Westlake Institute for Advanced Study, Hangzhou 310024, P. R. China}
\author{Zhuokai Xu}
\affiliation{Key Laboratory for Quantum Materials of Zhejiang Province, Department of Physics, School of Science and Research Center for Industries of the Future, Westlake University, Hangzhou 310030, P. R. China}
\affiliation{Institute of Natural Sciences, Westlake Institute for Advanced Study, Hangzhou 310024, P. R. China}

\author{Jinjin Liu}
\affiliation{Centre for Quantum Physics, Key Laboratory of Advanced Optoelectronic Quantum Architecture and Measurement (MOE), School of Physics, Beijing Institute of Technology, Beijing 100081, China}
\affiliation{Beijing Key Lab of Nanophotonics and Ultrafine Optoelectronic Systems, Beijing Institute of Technology, Beijing 100081, China}
%\affiliation{Material Science Center, Yangtze Delta Region Academy of Beijing Institute of Technology, Jiaxing 314011, China}

\author{Ruiya Zhan}
%\affiliation{Key Laboratory for Quantum Materials of Zhejiang Province, Department of Physics, School of Science, Westlake University, Hangzhou 310030, P. R. China}
\affiliation{Key Laboratory for Quantum Materials of Zhejiang Province, Department of Physics, School of Science and Research Center for Industries of the Future, Westlake University, Hangzhou 310030, P. R. China}
\affiliation{Institute of Natural Sciences, Westlake Institute for Advanced Study, Hangzhou 310024, P. R. China}

\author{Zhiwei Wang}
%\email{zhiweiwang@bit.edu.cn}
\affiliation{Centre for Quantum Physics, Key Laboratory of Advanced Optoelectronic Quantum Architecture and Measurement (MOE), School of Physics, Beijing Institute of Technology, Beijing 100081, China}
\affiliation{Beijing Key Lab of Nanophotonics and Ultrafine Optoelectronic Systems, Beijing Institute of Technology, Beijing 100081, China}
\affiliation{Beijing Institute of Technology, Zhuhai 519000, China}

%\author{Yugui Yao}
%\affiliation{Centre for Quantum Physics, Key Laboratory of Advanced Optoelectronic Quantum Architecture and Measurement (MOE), School of Physics, Beijing Institute of Technology, Beijing 100081, China}
%\affiliation{Beijing Key Lab of Nanophotonics and Ultrafine Optoelectronic Systems, Beijing Institute of Technology, Beijing 100081, China}
%\affiliation{Material Science Center, Yangtze Delta Region Academy of Beijing Institute of Technology, Jiaxing 314011, China}

\author{Xiao Lin}
\email{linxiao@westlake.edu.cn}
%\affiliation{Key Laboratory for Quantum Materials of Zhejiang Province, Department of Physics, School of Science, Westlake University, Hangzhou 310030, P. R. China}
\affiliation{Key Laboratory for Quantum Materials of Zhejiang Province, Department of Physics, School of Science and Research Center for Industries of the Future, Westlake University, Hangzhou 310030, P. R. China}
\affiliation{Institute of Natural Sciences, Westlake Institute for Advanced Study, Hangzhou 310024, P. R. China}

\begin{abstract}
\noindent
\textbf{Superconducting chiral domains associated with a time-reversal symmetry-breaking order parameter have garnered significant attention in Kagome systems. In this work, we demonstrate both the intrinsic direct-current and alternating-current Josephson effects in the nanoplates of the vanadium-based Kagome material CsV$_3$Sb$_5$, as evidenced by Fraunhofer-like patterns and Shapiro steps. Moreover, both the Fraunhofer-like patterns and Shapiro steps are modulated by thermal cycling, suggesting that the Josephson effects arise from dynamic superconducting domains. These findings may provide new insights into chiral superconductivity in CsV$_3$Sb$_5$ and highlight the potential of these intrinsic Josephson junctions for applications in chiral superconductor-based quantum devices.}

%However, directly detecting these domains through general magnetization measurements remains challenging due to the subtle magnetic moments generated by the chiral domains. 
%\textbf{Keywords: }
\end{abstract}

\maketitle

%\section{Introduction}
%\vspace{-10 pt}

The Josephson effect in superconductor-normal metal-superconductor (SNS) junctions has been extensively studied due to its significant applications and rich, exotic physical phenomena~\cite{Barone1982book, RevModPhys.76.411, RevModPhys.73.357}. In these junctions, Cooper pairs across the normal metal without dissipation, resulting in Josephson supercurrent. Upon applying a perpendicular magnetic field ($B$), the Josephson critical current ($I_\textrm{c}$) displays a Fraunhofer-like oscillatory pattern as a function of magnetic flux ($\Phi$), representing the direct-current (d.c.) Josephson effect \cite{JJfirst}. Furthermore, when a d.c. voltage ($V_\textrm{dc}$) is applied across the Josephson junction (JJ), an alternating current (a.c) is generated, giving rise to the a.c. Josephson effect \cite{JJfirst}. Conversely, when the JJ is exposed to microwave irradiation, it produces quantized voltage steps with an interval of $hf/2e$, known as integer Shapiro steps \cite{PhysRevLett.11.80}, with $h$ being Plank’s constant, $f$ the microwave frequency and $e$  the electron charge.

In chiral superconductors, edge states arise at the boundaries between different superconducting (SC) domains, creating a scenario analogous to the normal metal layer in SNS junctions \cite{Bouhon_2010, PhysRevLett.104.147001, PhysRevB.64.214503, Weitering2023NP}. These edge states effectively connect neighboring domains, facilitating the occurrence of  intrinsic Josephson effects without the need for external junctions. This phenomenon is a natural consequence of the underlying chiral order parameter and may be observed spontaneously in simple chiral SC systems, although experimental observations of this effect are rare.

Kagome materials, particularly the vanadium-based family $A$V$_3$Sb$_5$ ($A$ = K, Rb, Cs), standing out as a notable example of exotic superconductivity~\cite{PhysRevB.85.144402,  PhysRevLett.127.177001, PhysRevB.106.174514, WangQH2013PRB, PhysRevLett.129.167001}. The crystal structure of these materials consists of a Kagome lattice, comprising a network of corner-sharing triangles and hexagons \cite{PhysRevMaterials.3.094407}, as illustrated in Fig. 1(a). They exhibit a unique combination of electronic features, including flat bands, van Hove singularities, Dirac points, unconventional charge-density-wave (CDW) phase, superconductivity and pair density wave (PDW), which together give rise to a variety of emerging physical properties \cite{kagomereview1, jiang2023kagome, yin2022topological, Zeljkovic2021nature, Miao2021PRX, Wilson2021PRX, Guguchia2022Nature, Hasan2021NM, Moll2022Nature, ChenXH2022Nature, Zeljkovic2022NP, Wilson2020PRL, GaoHJ2021Nature}.

Regarding the SC phase of $A$V$_3$Sb$_5$, extensive experimental studies~\cite{LuoJL2021CPL,Yuan2021SCPMA,PhysRevLett.127.187004,PhysRevB.104.174507,Shibauchi2023NC,Okazaki2023Nature}, including nuclear quadrupole resonance\cite{LuoJL2021CPL}, electron irradiation\cite{Shibauchi2023NC} and angle-resolved photoemission spectroscopy\cite{Okazaki2023Nature}, support the fully gapped s-wave pairing symmetry. While, 
%other observations~\cite{ChenXH2022nature1,hossain2025unconventional}, such as 
the competition between the SC and normal orders \cite{ChenXH2022nature1} and the residual quasiparticles in the SC state \cite{PhysRevLett.127.187004, GaoHJ2021Nature, ChenXH2021PRX, hossain2025unconventional, deng2024chiral}, suggest an unconventional nature. Furthermore, recent advances in experimental techniques have uncovered evidence of time-reversal symmetry-breaking SC domains, indicative of a chiral SC order parameter in these materials \cite{deng2024chiral, le2024superconducting, deng2024evidence, Le_2024, PhysRevB.111.014503}.

In this work, we report the observation of intrinsic Josephson effects in the SC state of CsV$_3$Sb$_5$ (CVS) nanoplates without the need for externally fabricated heterostructures. Fraunhofer-like patterns and Shapiro steps are observed using a four-terminal measurement configuration, providing clear evidence of the d.c. and a.c. Josephson effects, respectively. Remarkably, the intrinsic Josephson effects in CVS are modulated by thermal cycling, strongly suggesting the presence of dynamic SC domains with weak links at domain boundaries. This finding adds indispensable evidence supporting the expectation
of chiral superconductivity proposed in CVS.

\begin{figure}[h]
	%\begin{center}
		\includegraphics[width=0.49\textwidth]{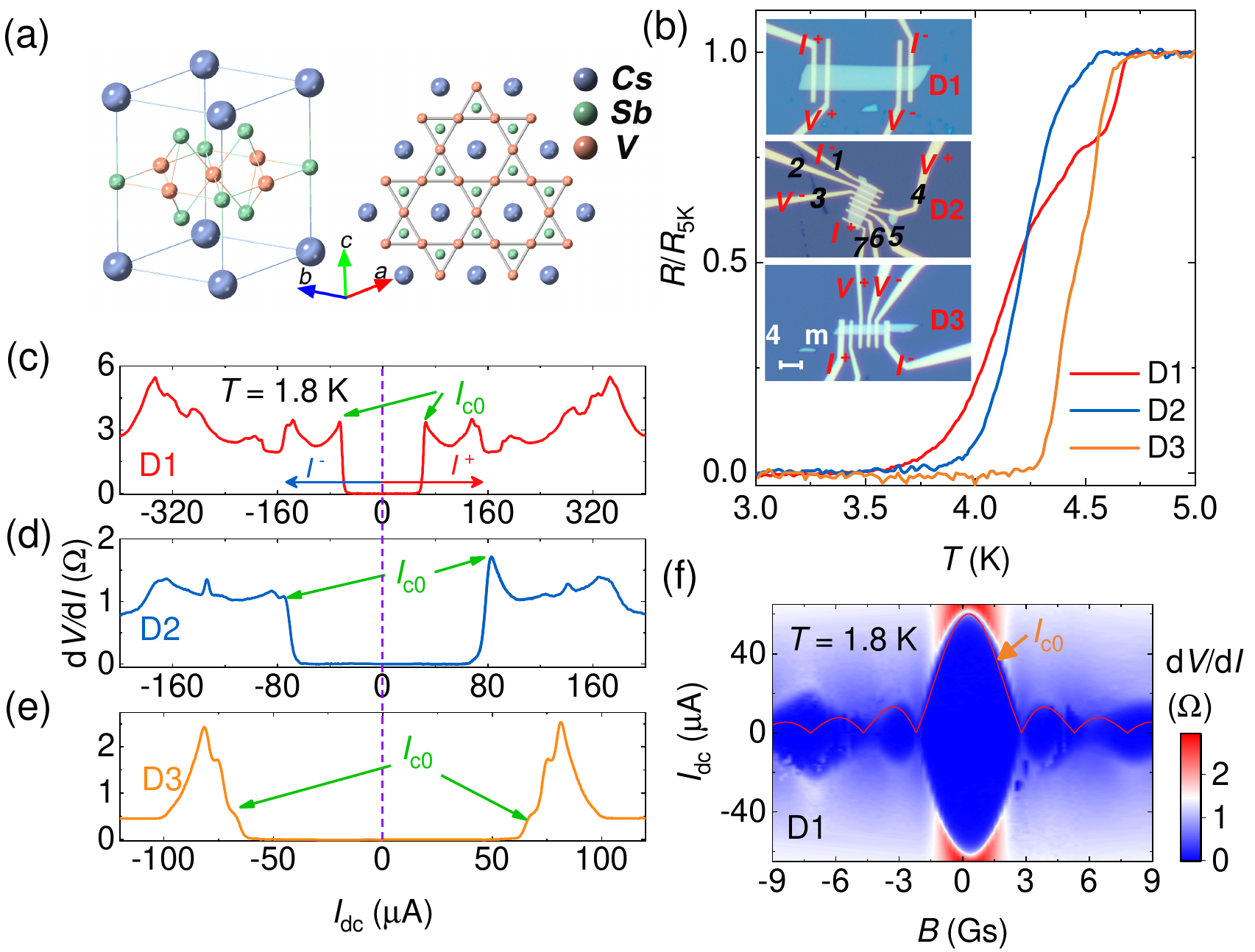}
	%\end{center}
	%\setlength{\abovecaptionskip}{-8 pt}
	\caption{(a) Crystal structure of CVS. (b) $R$ versus $T$ for D1, D2 and D3, respectively. The insets show photoimages of the devices. (c)-(e) d$V$/d$I$ versus $I_\textrm{dc}$ at 1.8 K for D1, D2 and D3. The red (blue) horizontal arrows indicate the sweep direction of bias current in the positive (negative) regime, i.e. $I^\textrm{+}$ ($I^\textrm{-}$). The green arrows mark the lowest critical supercurrent $I_\textrm{c0}$. (f) Color plot of d$V$/d$I$ as a function of $B$ and $I_\textrm{dc}$, measured at 1.8 K. $B$ is applied perpendicular to CVS nanoplates. The red curve is a simulation of the standard Fraunhofer pattern.}    
    %Magnetic field ($B$) dependence of the d$V$/d$I$-$I_\textrm{dc}$ curves at 1.8 K, with $B$ applied perpendicular to the CVS nanoplates. The red line represents a simulation based on the standard Fraunhofer pattern.	
	\label{Fig1}
\end{figure}

\begin{figure}[h]
	\begin{center}
		\includegraphics[width=0.49\textwidth]{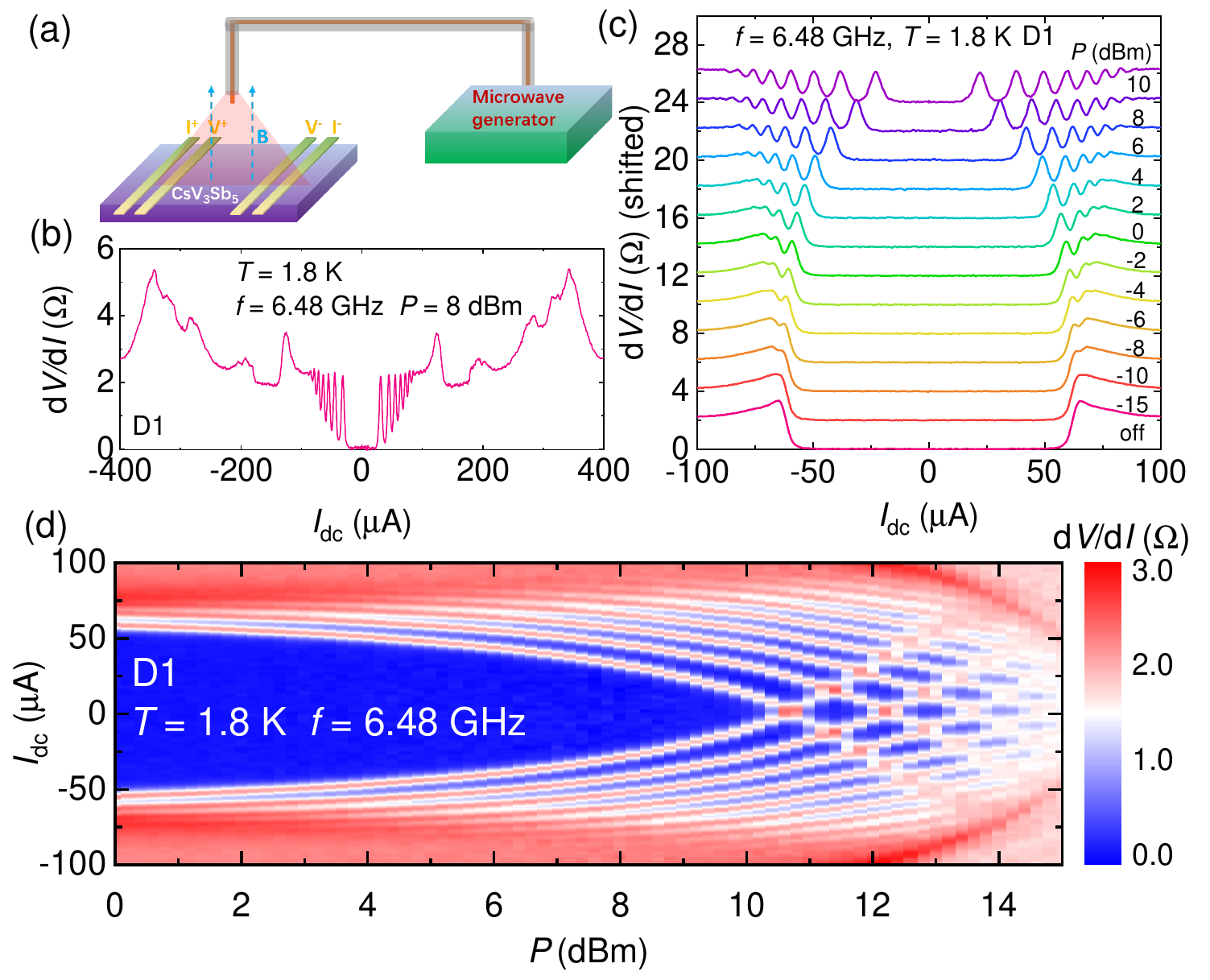}
	\end{center}
	\setlength{\abovecaptionskip}{-8 pt}
	\caption{(a) Schematic illustration of microwave irradiation on CVS nanoplates. (b) d$V$/d$I$ versus $I_\textrm{dc}$ for D1, irradiated with microwaves at a radio frequency $f$ = 6.48 GHz and power $P$ = 8 dBm at $T$ = 1.8 K. (c) d$V$/d$I$ versus $I_\textrm{dc}$ for D1 at various $P$ at $f$ = 6.48 GHz. (d) Corresponding d$V$/d$I$ maps as a function of $I_\textrm{dc}$ and $P$. 
    }
	\label{Fig2}
\end{figure}

Single crystals and devices of CVS (D1-D6) were grown and fabricated using the same methods as described in Ref.~\citenum{le2024superconducting}. The thickness of CVS nanoplates exceeds 30~nm.  The contacts were deposited with non-SC Ti (5~nm)/Au (80~nm) via electron beam evaporation. The measurements were performed by a standard four-terminal method in a Physical Property Measurement System (PPMS, Quantum Design) with additional experimental details provided in Supplementary Note I (See Supplementary Material\cite{SM, xiang2021twofold, PhysRevB.105.045102, PhysRevX.12.011001, PhysRevX.14.031015}).

Figure 1(b) illustrates the temperature ($T$) dependence of resistance ($R$) for D1, D2 and D3. All CVS nanoplates were exfoliated from the same high-quality single crystal with the residual resistivity ratio (RRR) of 250 \cite{le2024superconducting}. %These samples have been extensively distributed to research groups worldwide and thoroughly characterized via multiple experimental techniques \cite{le2024superconducting, Okazaki2023Nature, xiang2021twofold, PhysRevB.105.045102, PhysRevX.12.011001, PhysRevX.14.031015, hossain2025unconventional}. 
The SC transition temperature ($T_\textrm{c}^\textrm{0}$) identified as the zero-resistance point, varies between 3.5 K and 4.3 K. The transition width exhibits notable differences, likely associated with the property of SC domains. See more discussion in Fig.~S1 and Supplementary Note II in Supplementary Material\cite{SM}. %with D1 displaying two distinct transitions. Given that all devices were fabricated using identical techniques and the transition width can be thermally modulated (Fig. S1), external inhomogeneities or strain effects are unlikely to account for these variations. Instead, the distinctions in transition widths likely attributed to diverse configurations of superconducting domains within individual devices. 
To probe the critical supercurrent ($I_\textrm{c}$), Differential resistance (d$V$/d$I$) is examined as a function of d.c. bias current ($I_\textrm{dc}$) for D1, D2 and D3 in Figs. 1(c)-1(e), respectively. As $I_\textrm{dc}$ increases, multiple peaks emerge, exhibiting thermal modulation behavior in Fig. S1 (See Supplementary Material\cite{SM}), aligned with the expectation of SC domains~\cite{le2024superconducting}.

\begin{figure}[tb]
	\begin{center}
		\includegraphics[width=0.49\textwidth]{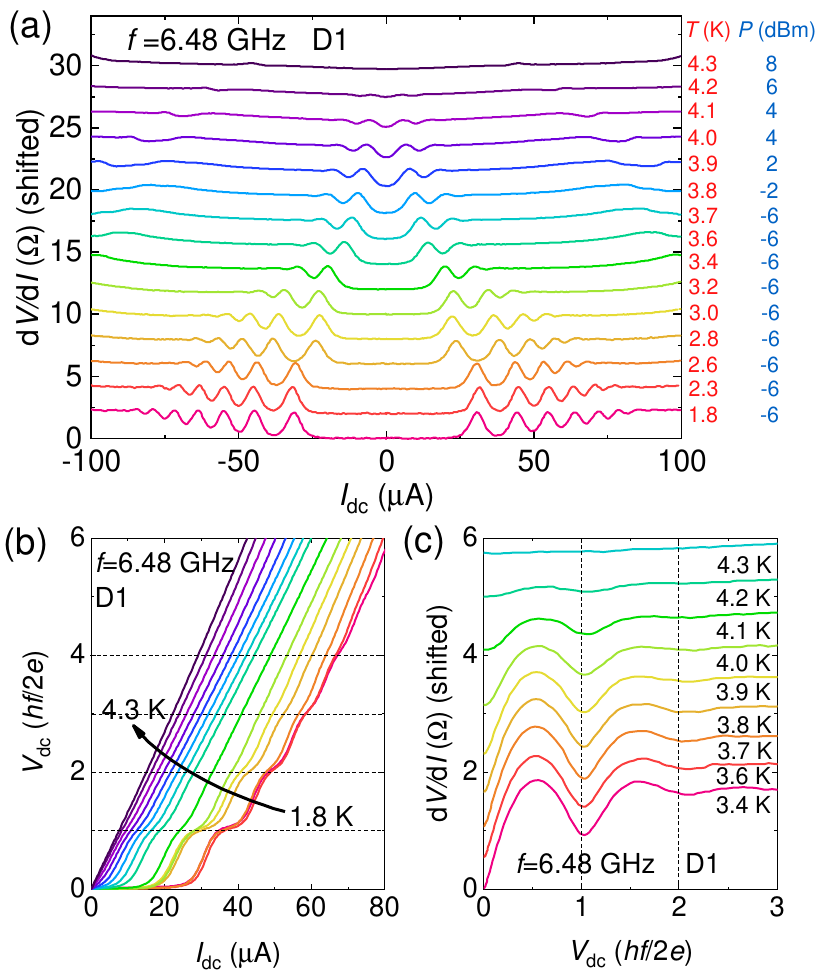}
	\end{center}
	\setlength{\abovecaptionskip}{-8 pt}
	\caption{(a) d$V$/d$I$ versus $I_\textrm{dc}$ for D1 at various $T$ and appropriate $P$ for $f$ = 6.48 GHz. (b) $V_\textrm{dc}$ versus $I_\textrm{dc}$ obtained by integrating the d$V$/d$I$-$I_\textrm{dc}$ curves in (a). $V_\textrm{dc}$ is normalized by the voltage $V_0 = hf/2e$. (c) d$V$/d$I$ as a function of normalized $V_\textrm{dc}$ at temperatures near $T_\textrm{c}^\textrm{0}$.
	}
    %to the integer Shapiro step separation voltage $V_0 = hf/2e$.
	\label{Fig3}
\end{figure}

Recent experiments in CVS, including phase-sensitive measurements\cite{le2024superconducting} and scanning tunneling microscopy \cite{deng2024chiral, deng2024evidence}, have reported signatures pointing to chiral superconductivity characteristic of time-reversal symmetry breaking and dynamic domain structures. In this context, one might expect Josephson effects to appear between neighbouring domains in certain configurations. To verify this hypothesis, d$V$/d$I$ - $I_\textrm{dc}$ curves were measured at selective $B$ on several devices to trace the behavior of $I_\textrm{c}$ subjected to $B$. 
Fraunhofer-like oscillations of $I_\textrm{c}$ were observed in D1, where $I_\textrm{c}$ is defined as the boundary of the blue region in the color map of Fig. 1(f). The standard Fraunhofer curve was simulated using the formula $I_\textrm{c}=I_0|sin(\pi \Phi /\Phi_0)/(\pi \Phi / \Phi_0 )|$, as represented by the red line in Fig. 1(f), where $I_0$ is the critical Josephson current at $B$ = 0 Gs and $\Phi_0=h/2e$ is the flux quantum. Notably, $I_\textrm{c}$ decays more gradually with increasing $|B|$ compared to the ideal Fraunhofer curve, a behavior offen observed in various types of JJs with the potential presence of boundary supercurrent \cite{le2024magnetic, kim2024edge, wang2021topology}. Similar Fraunhofer-like decay of $I_\textrm{c}$ was also identified in D3 and D5 in Fig. S2 (See Supplementary Material\cite{SM}), suggesting the existence of intrinsic DC Josephson effects in CVS. The effective JJ area, determined from the half-width of the central lobe in Fig. 1(f), is approximately 8.6 $\mu$m$^2$. This value is significantly smaller than the physical area enclosed by the current electrodes, as illustrated in the inset of Fig. 1(b). This observation supports the hypothesis that the JJ is formed between SC domains within the sample.

 %$\Phi$ is the magnetic flux, 

To further investigate the intrinsic Josephson properties, the devices were exposed to microwave irradiation at low $T$, as depicted in Fig. 2(a). In Fig. 2(b), additional d$V$/d$I$ peaks appear at $T$ = 1.8~K under microwave irradiation at a frequency $f$ = 6.48~GHz and power $P$ = 8~dBm, compared to those in Fig. 1(c). These additional d$V$/d$I$ peaks are induced by the microwave irradiation, evolving from the lowest critical current $I_\textrm{c0}$ as marked by the green arrow in Fig. 1(c). Figure 2(c) displays the evolution of the d$V$/d$I$-$I_\textrm{dc}$ curves for D1 by varing $P$ from the “off” state to 10 dBm. The multiple splittings of the $I_\textrm{c0}$ peak signal the appearance of Shapiro steps, which are characteristic of the a.c. Josephson effect \cite{JJfirst, PhysRevLett.11.80, Barone1982book}. The corresponding standard Shapiro map is illustrated in Fig. 2(d), constructed from the d$V$/d$I$ versus $I_\textrm{dc}$ and $P$ at $T$ =1.8 K, where the blue regions correspond to the separation between the d$V$/d$I$ peaks \cite{assouline2019spin, Yang_2019}. The observed a.c. Josephson effect further confirms the formation of intrinsic JJs in CVS nanoplates. Similar splitting of d$V$/d$I$ peaks under microwave irradiation is also reproduced in D3 and D5, as displayed in Fig. S3 and Fig. S4 (See Supplementary Material\cite{SM}).

Recently, the multiple electron paring in the vicinity of $T_\textrm{c}^\textrm{0}$ has attracted significant interest in CVS \cite{PhysRevX.14.021025, LI20242328, zhang2024higgs, PhysRevB.109.214509, zhou2022chern}. The a.c. Josephson effect serves as an effective probe for charge sensing, providing insights into the number of electrons forming a Cooper pair \cite{LI20242328, PhysRevResearch.6.033281}. In typical heterostructural JJs, the onset temperature of Josephson effects is generally well below $T_\textrm{c}^\textrm{0}$ owing to the  the imperfect nature  of the heterointerface, which restricts the detectable temperature range for paired electrons. In contrast, intrinsic JJs in CVS benefit from a high-quality interface between the SC regions and the edge states, allowing for a higher Josephson transition temperature. This enables the investigation of electron pairing behavior closer to $T_\textrm{c}^\textrm{0}$.

Figure 3(a) exhibits the $T$-evolution of d$V$/d$I$-$I_\textrm{dc}$ curves at several suitable microwave powers. The a.c. Josephson effect disappears around $T$ = 4.3 K, close to the temperature where the resistance starts to drop ($T_\textrm{c}^\textrm{onset}$). By integrating these d$V$/d$I$-$I_\textrm{dc}$ curves, the $V_\textrm{dc}$-$I_\textrm{dc}$ curves are obtained, with $V_\textrm{dc}$ normalized to the integer Shapiro step separation voltage $V_0 = hf/2e$, as shown in Fig. 3(b). Distinct integer Shapiro steps corresponding to indexes 1, 2, 3 and 4 are clearly visible, as indicated by the dashed lines in Fig. 3(b). To extract the indexes near $T_\textrm{c}^\textrm{0}$, d$V$/d$I$ versus normalized $V_\textrm{dc}$ ranging from 3.4 K to 4.3 K is displayed in Fig. 3(c), which distinctly demonstrates discrete Shapiro steps with an interval of integer quantized voltage. These results point to the presence of charge-2$e$ superconductivity at $T$ up to 4.3 K, corresponding to 0.91 times $T_\textrm{c}^\textrm{onset}$. It appears to contrast with charge 4$e$/6$e$ pairing revealed by previous Little-Parks measurements~\cite{PhysRevX.14.021025, LI20242328, zhang2024higgs, PhysRevB.109.214509, zhou2022chern}.
%{\color{red}While our observations are consistent with conventional charge-2$e$ pairing, alternative explanations involving charge 4$e$/6$e$ pairing cannot be definitively excluded at present \cite{PhysRevX.14.021025, LI20242328, zhang2024higgs, PhysRevB.109.214509, zhou2022chern}.}
Moreover, the standard integer Shapiro steps may imply the absence of topological SC order~\cite{assouline2019spin, rokhinson2012fractional}.
It's important to note that external confounding factors such as quasiparticle poisoning and heating effects, may obscure the intrinsic signals associated with all these intriguing phenomena \cite{dartiailh2021missing, le2019joule, PhysRevLett.118.137701}. Further detailed JJ-based experiments are required to elucidate these complex questions.

%which necessitating more detailed investigation in the context of a.c. Josephson effects. 

\begin{figure}[tb]
	\begin{center}
		\includegraphics[width=0.49\textwidth]{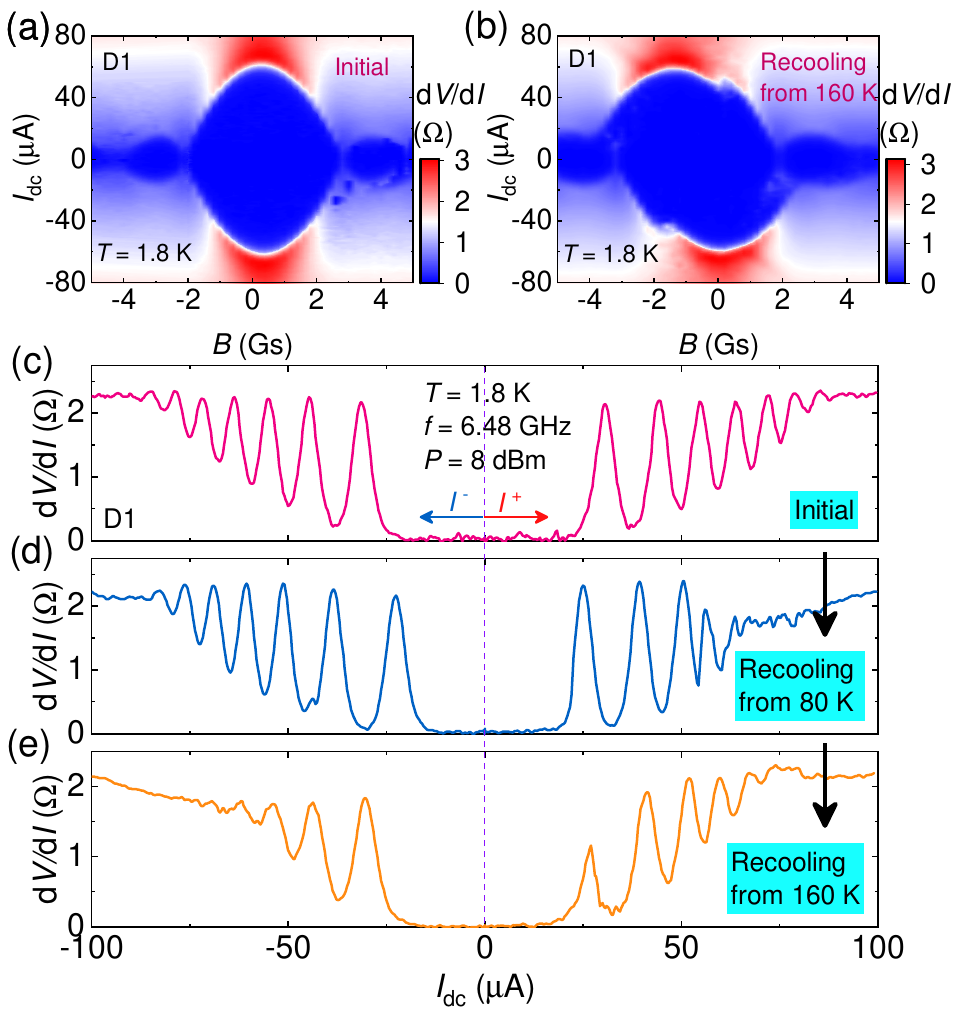}
	\end{center}
	\setlength{\abovecaptionskip}{-8 pt}
	\caption{(a) d$V$/d$I$ maps as a function of $I_\textrm{dc}$ and $B$ for the initial state of D1 at 1.8 K. (b) Repeated measurement for D1 after recooling from 160 K. (c) d$V$/d$I$ versus $I_\textrm{dc}$ under microwave irradiation at $f$ =6.48 GHz, $P$ = 8 dBm and $T$ = 1.8 K in the initial state. (d) and (e) Repeated measurement after recooling from 80 K and 160 K, respectively.
	}
	\label{Fig4}
\end{figure}

\begin{figure}[tb]
	\begin{center}
		\includegraphics[width=0.49\textwidth]{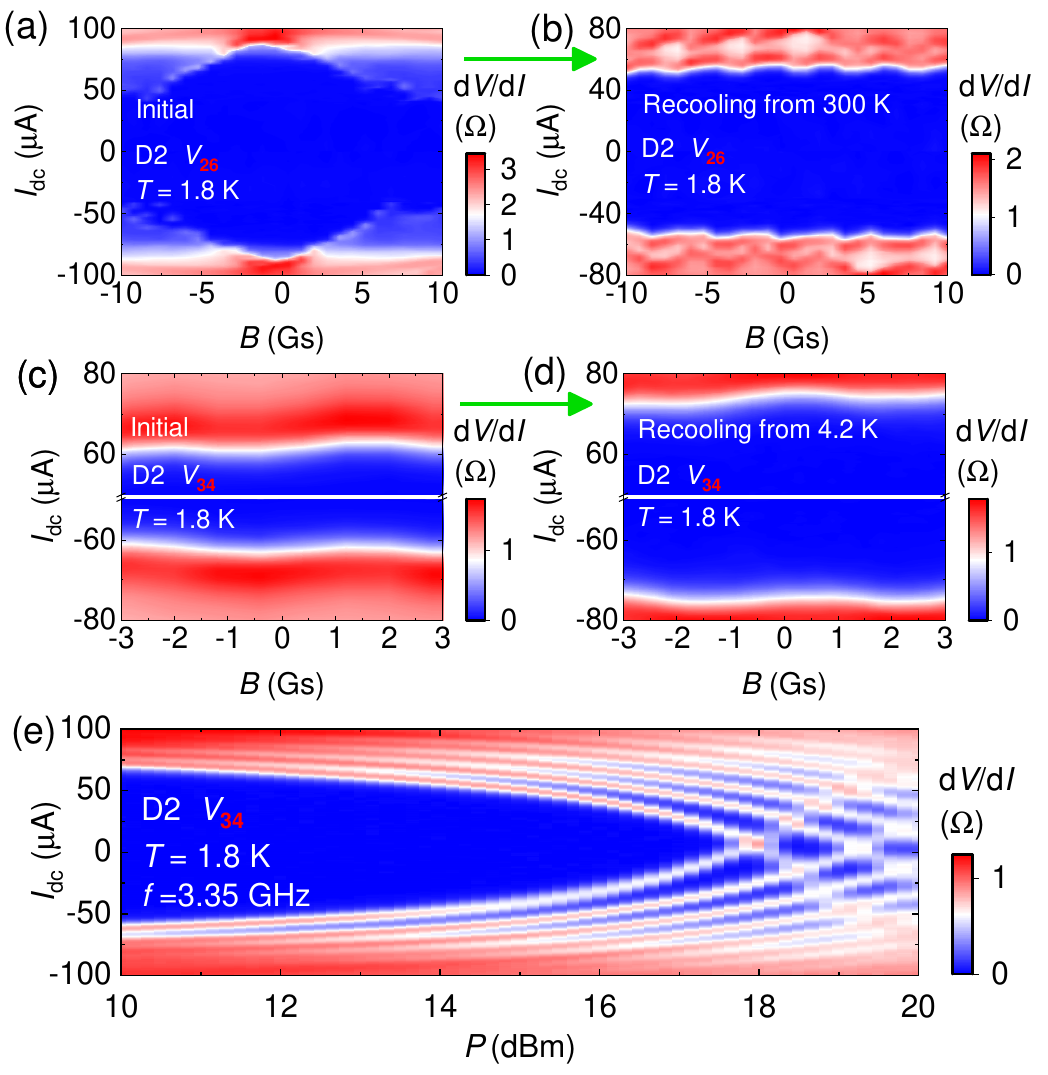}
	\end{center}
	\setlength{\abovecaptionskip}{-8 pt}
	\caption{(a) d$V$/d$I$ map as a function of $I_\textrm{dc}$ and $B$ for D2 on voltage terminals 2 and 6 ($V_{26}$) measured at 1.8 K, showing a Fraunhofer-like pattern. (b) Repeated measurement on $V_{26}$ after recooling from 300~K, showing significant modulation to a DSIP. (c) d$V$/d$I$ map for D2 on $V_{34}$ measured at 1.8 K. (d) Repeated measurement on $V_{34}$ after recooling from 4.2~K, showing apparment thermal modulation. (e) d$V$/d$I$ as a function of $I_\textrm{dc}$ and $P$ on $V_\textrm{3-4}$ at $f$ =3.35 GHz.}    
   % (a) and (c) d$V$/d$I$ maps as a function of $I_\textrm{dc}$ and $B$ for the initial state of D2 at 1.8 K. $V_\textrm{2-6}$ ($V_\textrm{3-4}$) represents the voltage measured between electrodes 2 (3) and 6 (4), as denoted in Fig. 1(a). (b) and (d) d$V$/d$I$ maps as a function of $I_\textrm{dc}$ and $B$ for $V_\textrm{2-6}$ ($V_\textrm{3-4}$) of D2 after recooling from 300 K (4.2 K). (e) d$V$/d$I$ as a function of $I_\textrm{dc}$ for $V_\textrm{3-4}$ of D2 at various $P$ at $f$ =3.35 GHz. 
	
	\label{Fig5}
\end{figure}

The observed intrinsic Josephson effect in CVS falls in the framework of chiral SC domains proposed in ref.~\citenum{le2024superconducting}. 
To provide further evidence, thermal modulation are performed to reveal the domain dynamics and exclude static inhomogeneities as the origin of the Josephson effect.
Figures 4(a) and (b) present d$V$/d$I$ versus $I_\textrm{dc}$ and $B$ mappings before and after recooling from 160 K, respectively. Notably, the width and symmetry of the central lobes are both altered by thermal cycling, implying a thermal modulation of the effective JJ areas and geometry. In addition, the quantity of observed Shapiro steps in both $I^+$ and $I^-$ regimes varies with the thermal cycling process, as observed after recooling from 80 K and 160 K, shown in Figs. 4(c)-(d). This may suggest a modulation of interface properties between adjacent domains. The pronounced SC diode effect, characterized by the asymmetry between the $I^+$ and $I^-$ regimes, observed in Figs. 4(b), (d), and (e), is in agreement with recent work \cite{le2024superconducting, PhysRevB.111.014503}. Figure S5 shows the thermal modulation of Josephson effects for D6 from $T\approx15$~K, well below the ordering temperature of the chiral CDW and nematic phases (See Supplementary Material\cite{SM}). It safely supports the role of SC domains and rules out the contribution of potential CDW or nematic domains~\cite{kagomereview1, jiang2023kagome} in generating dynamical Josephson effects. Further details are presented in Supplementary Note III (See Supplementary Material\cite{SM, guo2024correlated, Hardy2024PRL, Lai2005SM})

More intriguingly, Figs.~5(a) and (b) present a thermally modulated transformation from a Fraunhofer-like pattern to a double-slit superconducting interference pattern (DSIP, related to the loop supercurent structure)~\cite{le2024superconducting} for D2 on voltage electrodes $V_\textrm{2-6}$. This suggests a significant reorganization of SC domains after recooling from 300 K. Furthermore, in  Figs. 5(c) and 5(d), the DSIP on $V_\textrm{3-4}$ exhibits apparent thermal modulation even at $T\leq4.2$~K, which further diminishes the likelihood that normal-state domains contribute to the observation. On the other hand, strain was reported to have an impact on the electronic state of CVS~\cite{guo2024correlated}, leaving open possibility for the strain-induced Josephson phenomena. In our devices, strain might be introduced during thermal cycling due to differences in the thermal expansion coefficients between the device and the substrate. However, it should be emphasized that the strain induced by thermal cycling at low $T$, e.g. $T\leq4.2$~K, is negligibly small, which may not account for the thermal modulation in Fig. 5(c) and 5(d). Crucially, we detect well-defined Shapiro steps in D2 with the presence of DSIP, as displayed in Fig. 5(e). This indicates that the Josephson coupling is prevalent in CVS flakes, lending strong support to the framework of the chiral SC domain network with boundary supercurrent proposed in Ref.~\citenum{le2024superconducting}.

In the context of chiral domain scenario, external $B$ is anticipated to orient domains, thereby altering the SC properties in CVS. As depicted in Fig.~S6 (See Supplementary Material\cite{SM}), we observe subtle yet discernible differences in the $R-T$ curves under zero-field cooling (ZFC) and field cooling (FC) conditions. The reduced responsivity of these curves to the field modulation is discussed in Supplementary Note IV (See Supplementary Material\cite{SM}).

%Although our observations reveal the presence of chiral superconducting domains, the Josephson effects exhibit negligible differences between zero-field cooling (ZFC) and field cooling (FC) conditions. This behavior appears inconsistent with conventional chiral superconductor scenarios, where an external magnetic field is expected to eliminate domain walls, resulting in a single-domain state that would suppress Josephson effects. The observed insensitivity of the ZFC/FC measurements may arise from several factors, as discussed in the Supplementary Materials, including the persistence of multiple degenerate domains or weak domain-wall coercivity.}

Note that in layered superconductors, interlayer quantum tunneling of Cooper pairs may give rise to intrinsic Josephson effects, as observed in high $T_\textrm{c}$ cuprates~\cite{PhysRevB.49.1327, PhysRevLett.68.2394, chakravarty1993interlayer}. However, in our experiments, Fraunhofer-like patterns were measured with an in-plane current and out-of-plane $B$, showcasing in-plane JJs in CVS. The underlying mechanisms are distinct. In the former, the insulating blocks of layered cuprates acts as tunneling barriers, while in CVS, the Josephson barriers likely stem from weak links formed at the SC domain boundaries.

In conclusion, the thermally modulated intrinsic Josephson effects in CVS nanoplates have been examined. The appearance of Fraunhofer-like patterns and integer Shapiro steps unambiguously uncover the formation of Josephson junctions within the sample, which are tentatively attributed to weak links formed by edge states between chiral SC domains. A wealth of theoretical studies have predicted chiral superconductivity on the Kagome lattice~\cite{PhysRevB.85.144402, PhysRevLett.127.177001, PhysRevB.106.174514, WangQH2013PRB, PhysRevLett.129.167001, zhou2022chern}, with some theories suggesting a complex $d$$\pm$$id$ SC order parameter \cite{PhysRevB.85.144402, PhysRevLett.127.177001, PhysRevB.106.174514, WangQH2013PRB} and chiral PDW \cite{PhysRevLett.129.167001, zhou2022chern}in CVS. Recent experimental evidence for chiral Kagome superconductivity has emerged, supported by advanced techniques, such as phase-sensitive transport measurements and scanning tunneling microscopy \cite{le2024superconducting, deng2024chiral, deng2024evidence}. In this context, our findings of Josephson phenomena provide crucial evidence for the existence of SC domains within the chiral superconductivity framework, offering new insights into the SC nature of CVS. The intrinsic Josephson effects in CVS nanoplates also position them as promising candidates for quantum devices, with potential applications in advanced quantum technologies.

~\\
%{\color{red}$Data$ $availability$--The data that support the findings of this Letter are openly available\cite{Data}.}
~\\

%In conclusion, the thermally modulated intrinsic Josephson effects have been examined in CVS nanoplates. The appearance of Fraunhofer-like patterns and integer Shapiro steps unambiguously uncover the formation of Josephson junctions within the sample, which are tentatively attributed to weak links formed by edge states between chiral superconducting domains. These findings offer strong support for the chiral superconducting mechanism proposed in CVS and broaden the horizon for exploring chiral superconductors in quantum materials. The intrinsic Josephson effects in CVS nanoplates position them as promising candidates for quantum devices, with potential applications in advanced quantum technologies.

This research is supported by the National Key Research and Development Program of China under Grant No.2024YFA1408101 and National Natural Science Foundation of China under Grant No. 12474131. 
This research is supported by ``Pioneer'' and ``Leading Goose'' R$\&$D Program of Zhejiang under Grant 2024SDXHDX0007 and Zhejiang Provincial Natural Science Foundation of China for Distinguished Young Scholars under Grant No. LR23A040001. 
%This research is supported by the Research Center for Industries of the Future (RCIF) at Westlake University under Award No. WU2023C009.
T.L. acknowledges support from the China Postdoctoral Science Foundation (Grant No. 2022M722845 and No. 2023T160586).
Z.W. is supported by the National Key Research and Development Program of China (Grant Nos. 2022YFA1403400, 2020YFA0308800), Beijing National Laboratory for Condensed Matter Physics (Grant No. 2023BNLCMPKF007), and Beijing Natural Science Foundation (Grant No. Z210006). The authors thank the support provided by Dr. Chao Zhang from Instrumentation and Service Center for Physical Sciences at Westlake University.

\bibliographystyle{apsrev4-2}
%\noindent \textbf{References}
\bibliography{CVS_JJ}

\setcounter{figure}{0}

\renewcommand\thefigure{\textrm{S}\arabic{figure}}

\title{{\large Supplementary information} ~\\~\\ {Thermomodulated intrinsic Josephson effect in Kagome CsV$_3$Sb$_5$}}

\maketitle

\clearpage
\vspace{\baselineskip}
\noindent\textbf{\large{Supplementary Note I}}
\vspace{\baselineskip}

\noindent
The temperature ($T$) dependence of resistance ($R$) was measured using a standard four-terminal method in Quantum Design PPMS. A lock-in amplifier (SR830, Stanford Research) coupled with a 100 k$\Omega$ buffer resistor was used to apply a small alternating excitation current ($I_\textrm{ac}$) to detect the differential resistance (d$V$/d$I$). d$V$/d$I$ in the positive ($I^+$) and negative ($I^-$) bias regimes was measured by sweeping the d.c. current ($I_\textrm{dc}$) from 0 to $I_\textrm{max}^{+}$ or $I_\textrm{max}^{-}$. A current source meter (Keithley 2450) supplied $I_\textrm{dc}$. A current source meter (Keithley 2400) served as the power supply for the superconducting magnet, enabling precise control of the magnetic field ($B$) at sub-gauss levels. $B$ was applied along the $c$-axis, perpendicular to the large plane of the flakes. For the microwave irradiation experiments, a microwave analog signal generator (Sinolink Technologies, SLFS12A) was employed. Thermal cycling was conducted by using a 2 k$\Omega$ resistor as a heater, bonded to the SiO$_2$/Si substrate with silver paste. The local temperature of the devices was modulated by triggering the heater with a pulsed milliampere current supplied by a current source (Keithley 6221) for several seconds.

~\\
~\\
\vspace{\baselineskip}
\noindent\textbf{\large{Supplementary Note II}}

%\vspace{\baselineskip}
\begin{figure*}[thb]
	%\begin{center}
		\includegraphics[width=1\textwidth]{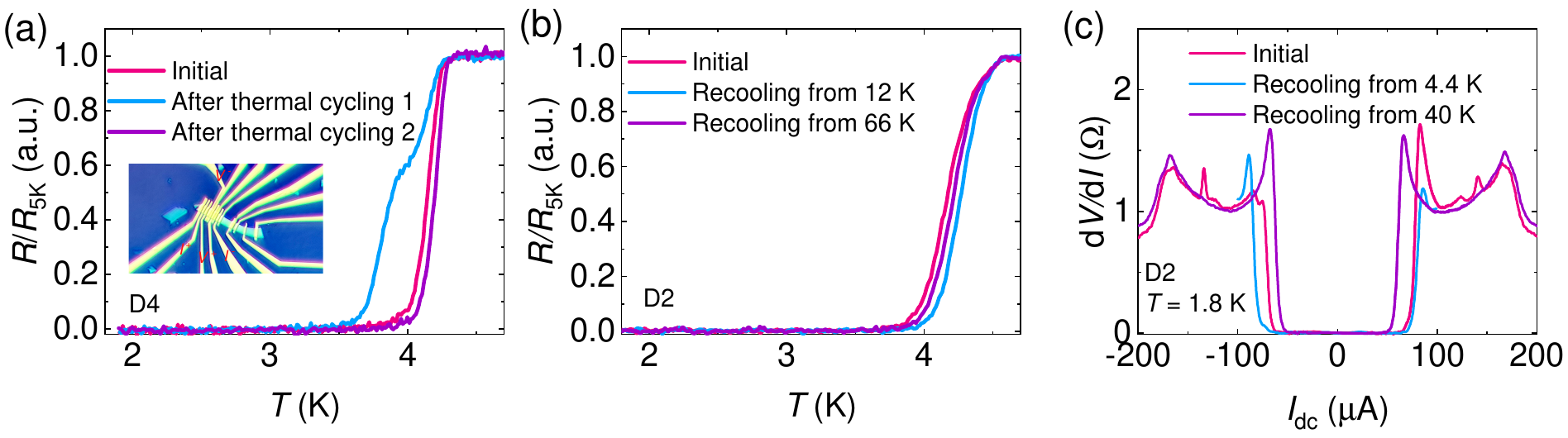}
	%\end{center}
	%\setlength{\abovecaptionskip}{-25 pt}
\caption{(a) Thermal modulation of $R$-$T$ curves at the superconducting transition for D4. The inset displays its optical microscope image as used in Ref. 39.
(b) Thermal modulation of $R$-$T$ curves at the superconducting transition on another device D2. The superconducting transition width is modulated by recooling the device from both 12 K and 66 K, respectively. (c) Thermal modulation of d$V$/d$I$-$I_\textrm{dc}$ curves by recooling from 4.4 K and 40 K, respectively.
}	
	\label{Fig.S1}
\end{figure*}

\noindent
In Fig. 1(b), the superconducting transition width of D1 is wider than that of the other devices. Below, we argue that this phenomenon is unlikely ascribed to significant sample inhomogeneity. Instead, it is might stem from the intrinsic characteristics related to the superconducting domains.

First of all, the residual resistivity ratio of our bulk crystal is about 250, a value among the highest ever reported, which indicates the ultra-high quality of the single crystals. Similar samples from the same group have been extensively distributed to research groups worldwide and thoroughly characterized via multiple experimental techniques \cite{le2024superconducting, Okazaki2023Nature, xiang2021twofold, PhysRevB.105.045102, PhysRevX.12.011001, PhysRevX.14.031015, hossain2025unconventional}, such as transport measurements, scanning tunneling microscopy, angle-resolved photoemission spectroscopy, and muon spin rotation. Given this extensive validation, the likelihood of bulk sample inhomogeneities is exceedingly low. 

Second, we have conducted extensive thermal modulation measurements of the $R-T$ curves for different devices as shown in Fig. S1. In Fig. S1(a), for D4, the superconducting transition width is notably affected by thermal cycling. As seen in the figure, the onset critical temperature ($T_\mathrm{c}^\mathrm{onset}$) is about 4.2~K  for all three curves. In the initial curve (pink), the device exhibits a sharp superconducting transition with $T_\mathrm{c}^{0}$ of 4~K. After the first thermal cycling (blue), the transition remarkably broadens, lowering $T_\mathrm{c}^{0}$ to 3.6~K. Following additional thermal cycling, the $R$-$T$ curve (magenta) returns to its initial state. 
Similarly, in Fig. S1(b), for D2, comparable results are obtained when the device is recooled from 12 K and 66 K, respectively. Additionally, as shown in Fig. S1(c), the critical current ($I_\textrm{c}$) of D2 is also modified upon recooling the device from 4.4 K and 40 K, respectively. Collectively, the thermal modulation of the superconducting transition width and critical current, particularly at temperatures (4.4~K or 12~K) well below the ordering temperature of the normal-state CDW and nematic phases,  strongly indicates the presence of dynamic superconducting domains influencing the superconducting properties of CVS, rather than hypothetic static inhomogeneities.

Generally, a broader superconducting transition signals the presence of significant weak links, which can arise from regions of static inhomogeneities or, as in our study, domain boundaries intrinsic to superconducting order.  Intuitively, the more prominent these weak links, the more noticeable the Josephson effect becomes. This principle aligns precisely with our experimental observations. D1, distinguished by its broader transition, exhibits a distinct Fraunhofer pattern in Fig.~1(f), enabling us to extract the paired electron charge from Shapiro step measurements in Fig.~3. In contrast, other devices with relatively sharper transition widths showed a smeared Fraunhofer pattern, as seen in Fig.~5 and Fig.~S2. However, the clear observation of Shapiro steps in these samples (Fig.~5(e) and Fig~S4)
demonstrates that the a.c. Josephson effect remains robust, insensitive to  the strengh of weak links. All these observations provide inportant evidence for the formation of intrinsic Josephson junctions at superconducting domain boundaries in all devices.

\begin{figure*}[thb]
  \includegraphics[width=1\textwidth]{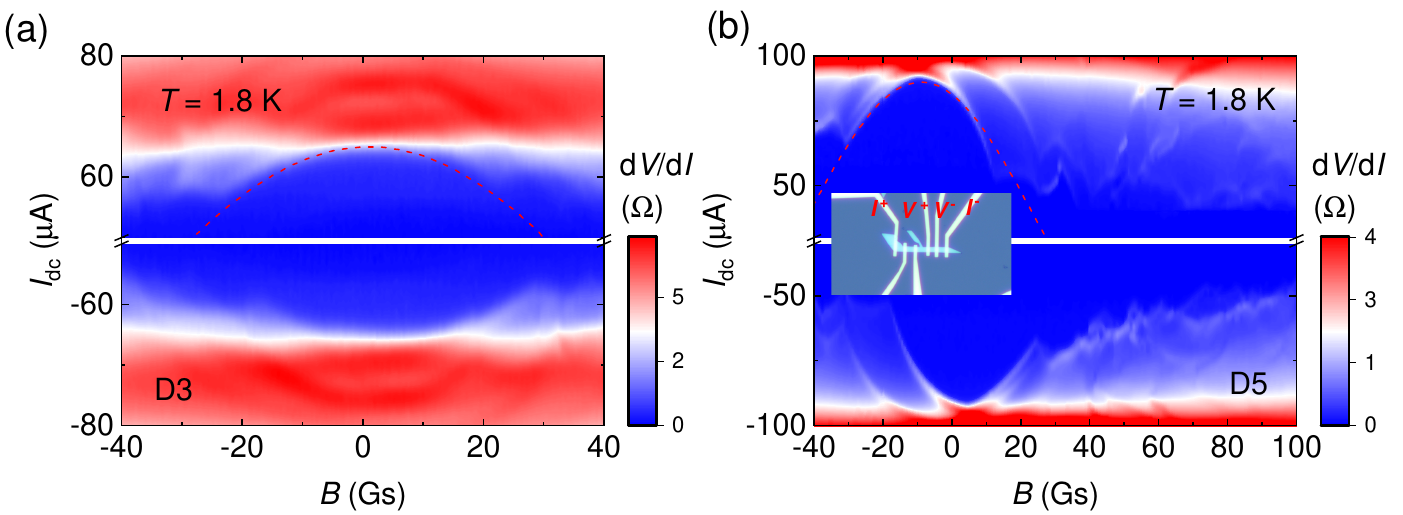}
   \caption{(a,b) Color plots of d$V$/d$I$ as a function of $B$ and $I_\textrm{dc}$, measured at 1.8 K for D3 and D5, respectively. $B$ is applied perpendicular to the CVS nanoplates. The red curve represents a simulation of the standard Fraunhofer pattern. The inset of (b) shows the photo image of D5.
	}
	\label{FigS2}
\end{figure*}

\begin{figure*}[thb]
	\includegraphics[width=1\textwidth]{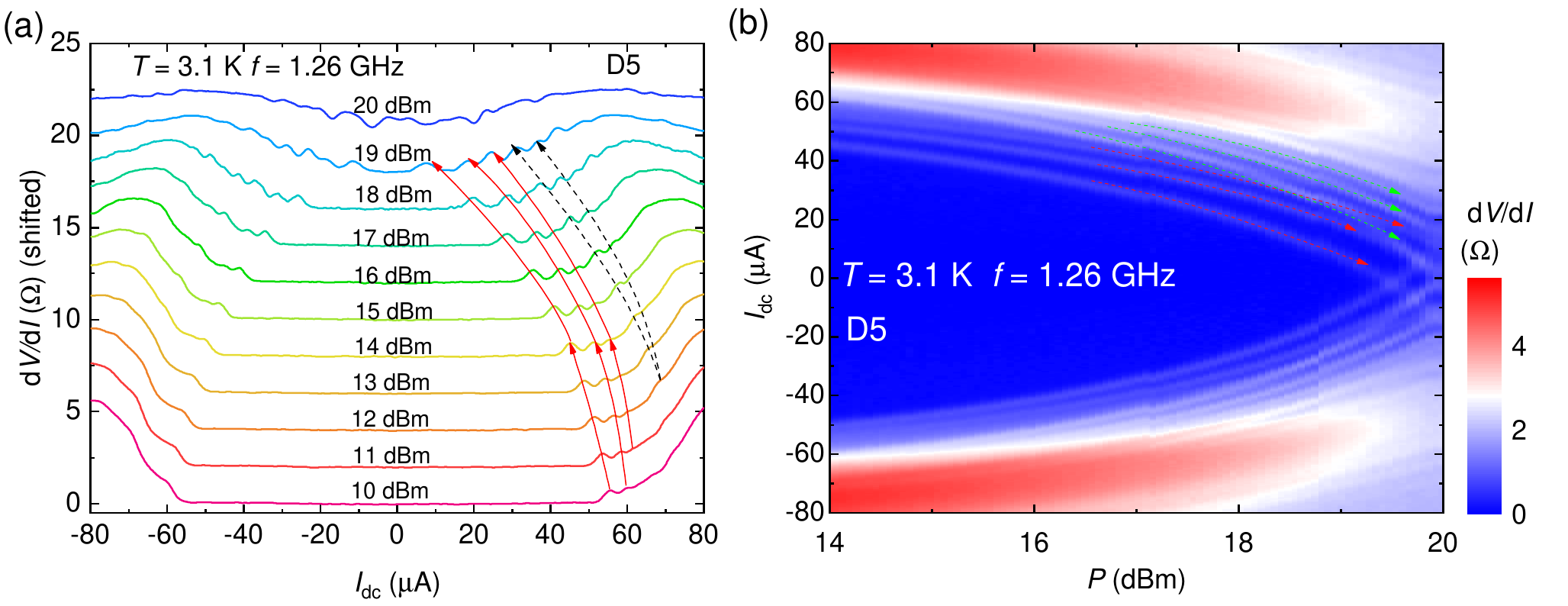}
	\caption{(a) d$V$/d$I$ as a function of $I_\textrm{dc}$ for D5 irradiated by microwaves at a radio frequency $f$ = 1.26~GHz and various microwave powers ($P$). The red and black dashed lines denote the splitting of two d$V$/d$I$ peaks under the microwave irradiation. (b) Corresponding d$V$/d$I$ maps as a function of $I_\textrm{dc}$ and $P$. The red and green dashed lines mark two sets of Shapiro steps that arises from two different JJs in D5.
    	}
	\label{FigS3}
\end{figure*}

\begin{figure*}[thb]
	\includegraphics[width=0.45\textwidth]{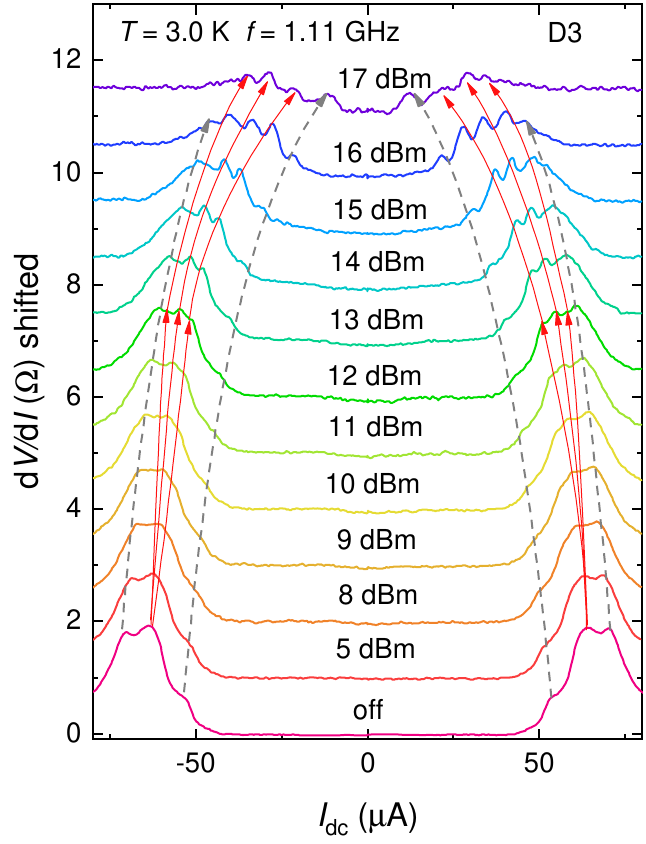}
	\caption{d$V$/d$I$ as a function of $I_\textrm{dc}$ for D3 irradiated by microwaves at $f$ = 1.11~GHz and various $P$.
    The red lines denote the splitting of d$V$/d$I$ peaks under the microwave irradiation. The black dashed lines denote the insensitivity of other d$V$/d$I$ peaks to microwave irradiation.
	}
	\label{FigS4}
\end{figure*}

\clearpage

\vspace{\baselineskip}
\noindent\textbf{\large{Supplementary Note III}}
\vspace{\baselineskip}

\begin{figure*}[thb]
		\includegraphics[width=0.9\textwidth]{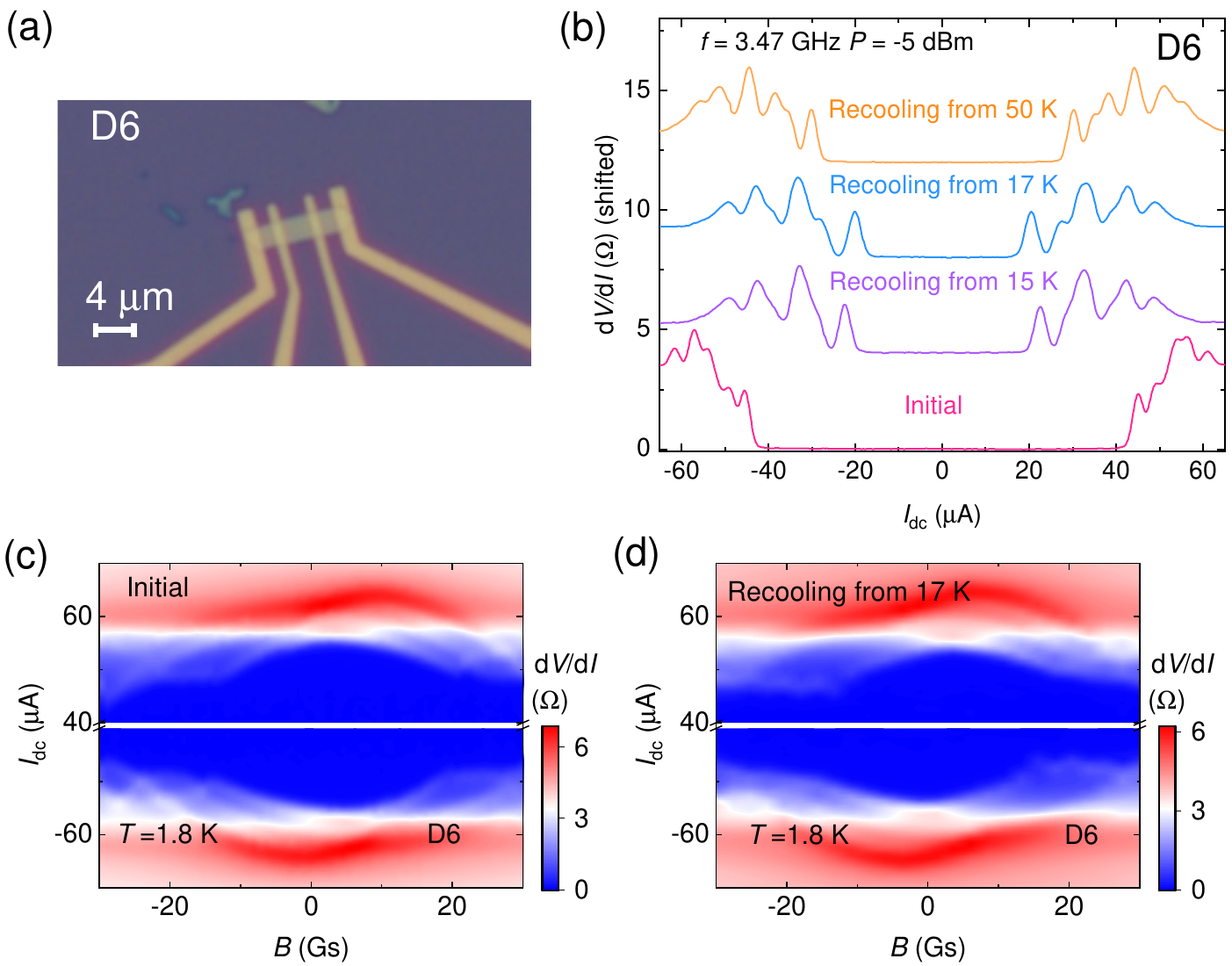}

	\caption{(a) Optical microscope image of D6. (b) d$V$/d$I$ as a function of $I_\textrm{dc}$ under microwave irradiation at $f$ =3.47 GHz, $P$ = -5 dBm and $T$ = 1.8 K, both in the initial state and after multiple thermal cycling. (c) d$V$/d$I$ maps as a function of $I_\textrm{dc}$ and $B$ for the initial state at 1.8 K. (d) Altered color map from repeated measurements at 1.8 K after recooling from 17 K, well below the CDW transition temperature. The modulation at low $T$ indicates the minimal influence of CDW dynamics on out results in the superconducting state.  
	}
	\label{FigS5}
\end{figure*}

\noindent
The ordering temperatures of the CDW and nematic orders in CVS are approximately 90~K and 30~K, respectively~\cite{kagomereview1, jiang2023kagome}. These two orders are expected to form domains in the normal state, which leaves open possibility for the observation of Josephson effects in our devices, although the underlying mechanism is yet to be solved. To exclude the possible origin from normal state domains, thermal cycling measurements conducted below the respective ordering temperatures are required. In Fig.~5, we present the thermal modulation of both a.c. and d.c. Josephson effect from sufficiently low temperatures for D6. Figure~S5(b) reveals that microwave irradiated d$V$/d$I$-$I_\textrm{dc}$ curves exhibit modulation upon recooling from 15 K or 17 K, reflecting dynamic Josephson coupling. Furthermore, Figs.~S5(c) and (d) showcase the thermal modulation of Fraunhofer-like patterns, which become more asymmetric between the positive and negative branches after recooling from 17 K. Collectively, these observations points to the role of superconducting domains in generating the intrinsic Josephson effects, instead of normal state domains. 

It has been reported that the strain has influence on the normal-state properties of CVS~\cite{guo2024correlated}. Therefore, it is reasonable to expect the impact of strain on the superconducting phase. In our case, a potential source of strain is the mismatch in thermal expansion coefficients between the CVS flake and the Si/SiO$_2$ substrate. To mitigate this, we intentionally performed thermal cycling at low temperatures, a procedure designed to minimize the strain effect. Below, we provide an upper-bound estimation of this thermally induced strain to demonstrate that it cannot account for the thermal modulation phenomena. Based on the values of the literature, the in-plane thermal expansion coefficient of CVS at low temperatures is $\alpha_\mathrm{CVS}\approx 10^{-7}~\mathrm{K}^{-1}$~\cite{Hardy2024PRL}; the thermal expansion coefficient of SiO$_2$ films is  $\alpha_\mathrm{SiO_2}\approx 2.4\times10^{-7}~\mathrm{K}^{-1}$. This value is measured at room temperature and is expected to significantly lower at cryogenic temperatures, thus serving as a conservative upper bound~\cite{Lai2005SM}. Assuming a rigid bond between the flake and substrate, the maximum differential strain generated upon cooling from 12~K to 2~K is estimated as:

\begin{equation}
		\varepsilon\approx(\alpha_\mathrm{CVS}-\alpha_\mathrm{SiO_2})\times\Delta T\approx1.4\times10^{-4}	
\end{equation}

\noindent
Note that this value represents a significant overestimation, as the flakes were mechanically transferred, resulting in a weak bond. Previous studies have shown that a strain on the order of $0.1\%$ is necessary to observably alter the superconducting properties of CVS~\cite{NiN2021PRB}. As our estimated strain is approximately three orders of magnitude smaller, we safely argue that the thermally induced strain is too small to account for the thermal modulation effects observed.

\clearpage

\vspace{\baselineskip}
\noindent\textbf{\large{Supplementary Note IV}}
\vspace{\baselineskip}
\begin{figure*}[thb]
		\includegraphics[width=0.5\textwidth]{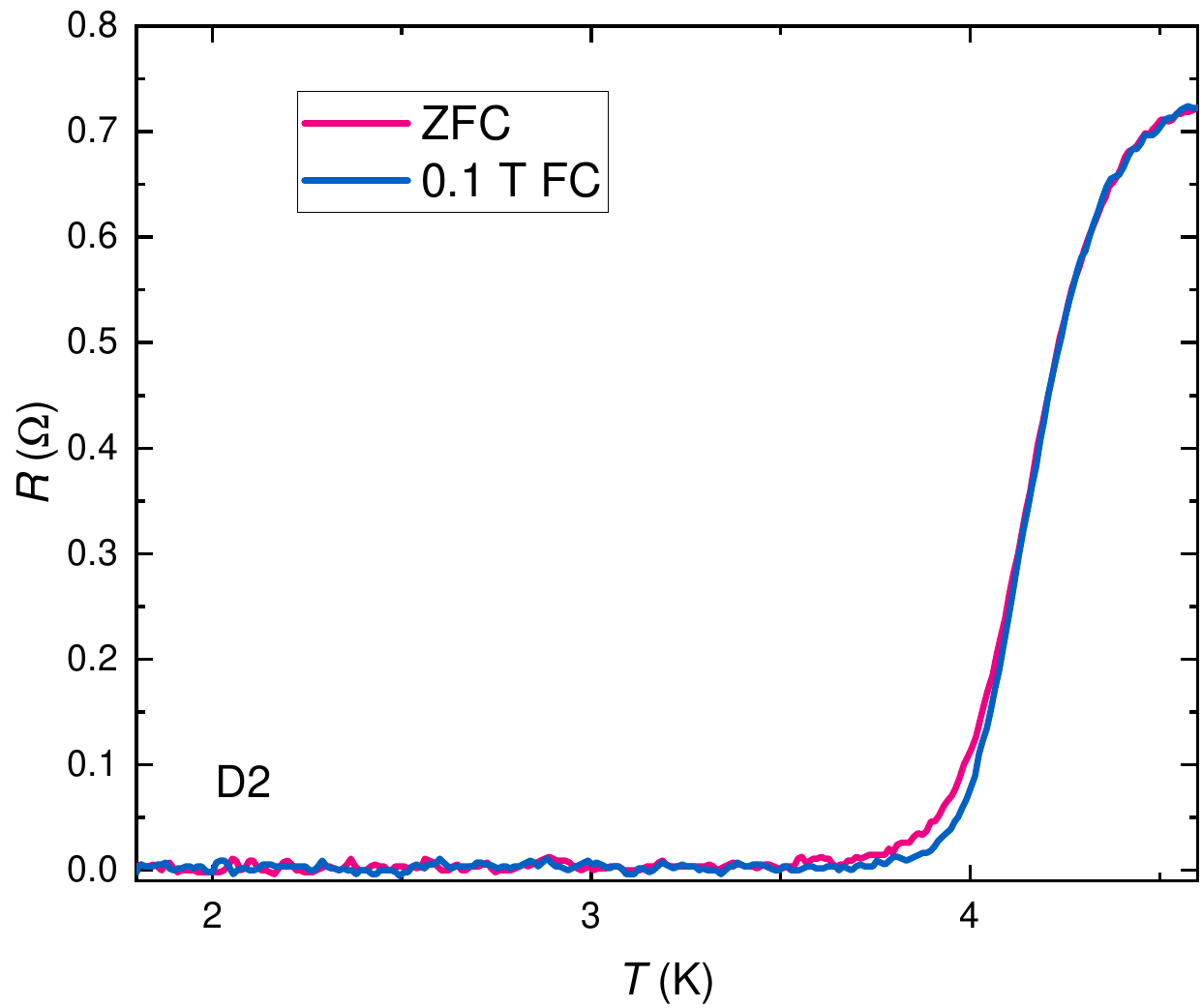}
	\caption{ZFC/FC measurements of $R-T$ curves on new device D2.}
	\label{FigS6}
\end{figure*}

\noindent
We have performed extensive zero-field cooling/field cooling (ZFC/FC) measurements on multiple devices. As illustrated by Fig.~S6 for D2, the measurements revealed a marginal yet detectable reduction in the superconducting transition width under FC conditions, with the Josephson effects remaining observable. We discuss below two possible explanations for the low sensitivity of R-T curves to the field modulation. 

First of all, there are three sublattices in Kagome systems. Theoretically, the chiral superconducting order parameters differ in phase by $2\pi/3$ between each sublattice. Each of these orders accommodates two degenerate phases with opposite chirality. In CVS, the $C_2$ rotation symmetry further disrupts the balance among the sublattices. All this leads to superconducting domains composed of multiple chiral superconducting orders. This complex domain structure is likely to render the system less responsive to external magnetic field. The existence of multiple superconducting domains remains an open question. Further experimental investigations are essential to elucidate this aspect in the future. 

Second, another possible explanation lies in the relatively weak coercivity of chiral superconducting domains, as revealed by the weak hysteresis of magnetoresistance reported in previous studies~\cite{Le_2024, PhysRevB.111.014503}. It may allow the system to recover multi-domain structures when the external field returns to zero.

\vspace{\baselineskip}

%\bibliographystyle{apsrev4-2}
%\noindent \textbf{References}
%\bibliography{CVS_JJ}

\end{document}